\documentclass[a4paper]{article}
\usepackage{amsfonts,amsmath,subfigure,cite,epsfig}
\usepackage{multirow}
\usepackage{lipsum}
\usepackage{INTERSPEECH2020}
\begin{document}
\title{Environmental Sound Classification with Parallel Temporal-spectral Attention}
\name{Helin Wang$^1$, Yuexian Zou$^{1,2,*}$, Dading Chong$^1$, Wenwu Wang$^3$ \thanks{*Corresponding author: zouyx@pku.edu.cn}}
\address{
  $^1$ADSPLAB, School of ECE, Peking University, Shenzhen, China\\
  $^2$Peng Cheng Laboratory, Shenzhen, China\\
  $^3$Center for Vision, Speech and Signal Processing, University of Surrey, UK}
\email{\{wanghl15,zouyx,1601213984\}@pku.edu.cn, w.wang@surrey.ac.uk}

\maketitle

\begin{abstract}
  Convolutional neural networks (CNN) are one of the best-performing neural network architectures for environmental sound classification (ESC). 
  Recently, temporal attention mechanisms have been used in CNN to capture the useful information from the relevant time frames for audio classification,
  especially for weakly labelled data where the onset and offset times of the sound events are not applied. 
  In these methods, however, the inherent spectral characteristics and variations are not explicitly exploited when obtaining the deep features. 
  In this paper, we propose a novel parallel temporal-spectral attention mechanism for CNN to learn discriminative sound representations, which enhances the temporal and spectral features by capturing the importance of different time frames and frequency bands.
  Parallel branches are constructed to allow temporal attention and spectral attention to be applied respectively in order to mitigate interference from the segments without the presence of sound events. 
  The experiments on three environmental sound classification (ESC) datasets and two acoustic scene classification (ASC) datasets show that our method improves the classification performance and also exhibits robustness to noise.
  
\end{abstract}

\section{Introduction}

Environmental sound classification (ESC) is an important research area in human-computer interaction aiming to classify an environment by its ambient sound,
 with a variety of potential applications such as audio surveillance \cite{radhakrishnan2005audio} and smart room monitoring \cite{vacher2007sound}. 
Due to the dynamic and unstructured nature of acoustic environments, it is a practical challenge to design appropriate features for environmental sound classification. 
In many existing ESC methods, the features are often designed based on prior knowledge of acoustic environments, and a classifier is then trained with the features to obtain the category probability of each environmental sound signal.

Among these methods, deep learning, which is facilitated by the availability of increased amount of training data and techniques of data augmentation, has been widely used in ESC.
Convolutional neural networks (CNN) based methods \cite{piczak2015environmental,lee2017raw,huzaifah2017comparison,zhang2017dilated,zhang2018deep,weiping2017acoustic} offer the state-of-the-art performance, where spectrograms and mel-scale frequency cepstral coefficients (MFCC) are often used as the input of the networks. 
Different from the images in visual recognition tasks, however, the temporal and spectral information represented by spectrograms will have different characteristics and degree of importance in sound recognition. 
Although the translation of local patterns in the time domain has little effect on the classification of sound events, the difference across frequency bands has a significant impact on the performance of sound classification \cite{koutini2019receptive}. 
To capture the information about which parts of the features are more relevant to the sound events, attention mechanisms have been proposed \cite{xu2017attention,kong2019weakly,zhang2019attention,ren2018attention,kong2018audio,wang2019acoustic,phan2019spatio,hong2020weakly}, 
especially for weakly labelled data where the timing information about the sound events is not available in the training data. 
In these methods, temporal attention \cite{kong2019weakly,kong2018audio} is applied to obtain weights for combining feature vectors at different time steps, however, the importance of different frequency bands is not considered. 
Spatial attention \cite{hong2020weakly} characterizes the importance of the regions with spatial weights to target the location of sound events, but ignores the inherent time-frequency characteristics. 

To address the above issues, we propose a parallel temporal-spectral attention mechanism for CNN to learn discriminative time-frequency representations, which allows the networks to be aware of the variety of information in time frames and frequency channels.
Specifically, temporal attention is employed to capture the certain frames where sound events appear, and a spectral attention method is proposed to pay a different degree of attention to various frequency bands. 
The idea of the spectral attention is inspired by the study of frequency-selective attentional filter in human primary auditory cortex \cite{da2013tuning}, which shows that the human brain facilitates selective listening to a frequency of interest in a scene by reinforcing the fine-grained activity pattern throughout the entire superior temporal cortex that would be evoked. 
In addition, the parallel structure is applied in our method, which mitigates interference between the temporal and spectral features by paying attention with two different branches,
and also promotes the robustness when a single branch is disturbed by the segments without the presence of sound events.
The proposed method is evaluated on three benchmark datasets (ESC-10 \cite{piczak2015esc}, ESC-50 \cite{piczak2015esc} and UrbanSound8k \cite{salamon2014dataset}),
and achieves the state-of-the-art classification accuracy of $95.8\%$, $88.6\%$ and $88.5\%$, respectively.
Furthermore, our method is applied to another audio classification task, \textit{i.e.} acoustic scene classification (ASC), and also improves the performance.

\section{Proposed Method}
In this section, the temporal attention and spectral attention methods are introduced, which enhance the features from relevant time frames and frequency bands.
In order to obtain the temporal and spectral features simultaneously, a parallel temporal-spectral attention mechanism is then introduced.
\begin{figure}[t] 
   \centering
   \subfigure[Temporal attention]{\label{h1}
   \begin{minipage}[c]{0.5\textwidth}
   \centering
   \includegraphics[width=2.8in]{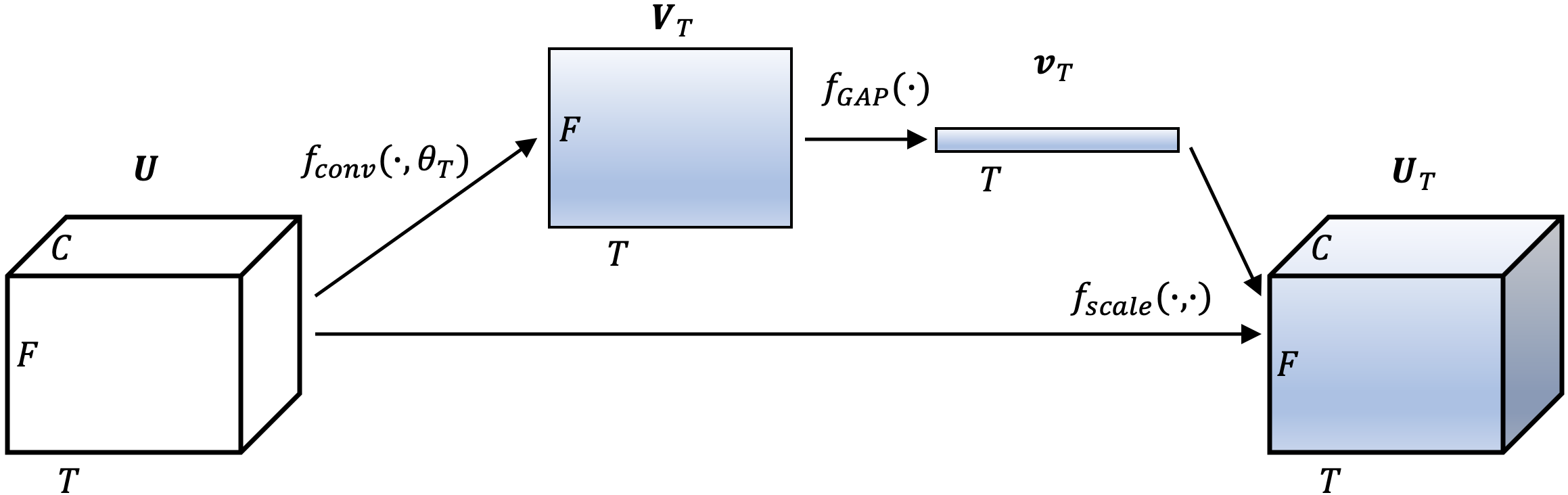}
   \end{minipage}
   }
   \subfigure[Spectral attention]{\label{h2}
   \begin{minipage}[c]{0.5\textwidth}
   \centering
   \includegraphics[width=2.8in]{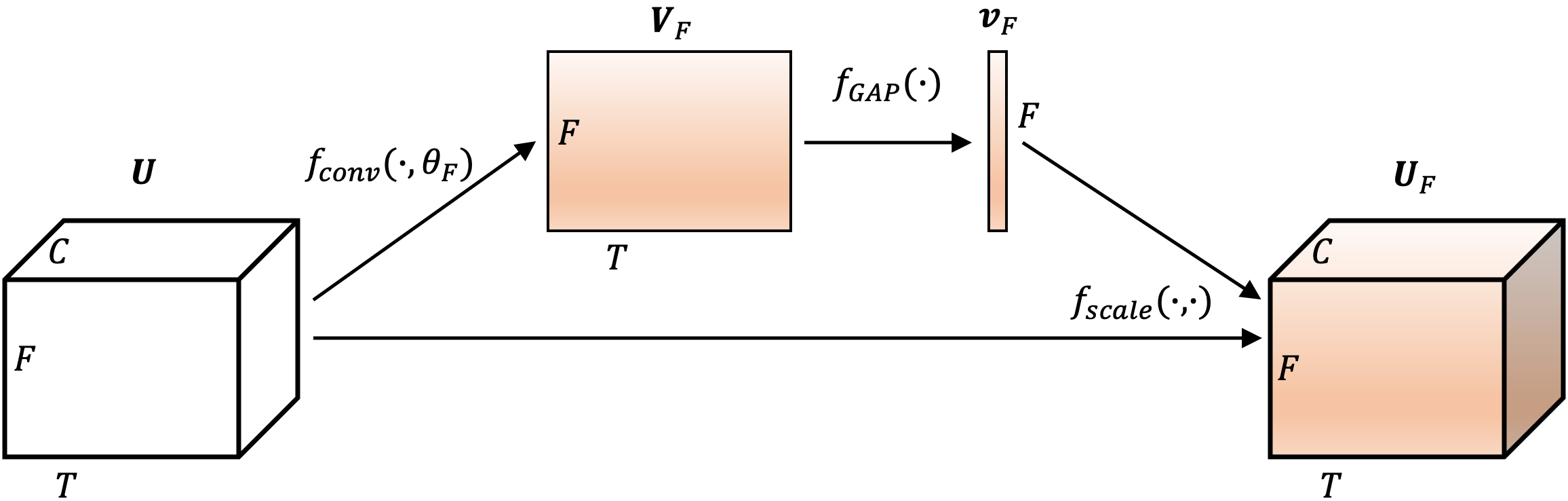}
   \end{minipage}
   }
   \caption{The illustration of temporal attention and spectral attention mechanisms.}
   \label{fig1}
\end{figure}

\subsection{Temporal Attention and Spectral Attention}
CNN have been widely used in the audio classification tasks, which show powerful ability to extract high-level features from low-level features, such as log mel spectrogram.
To start with an input spectrogram of size $T \times F \times 1$, 
the convolutional layer consisting of $C$-channel filters outputs a $T^{\prime} \times F^{\prime} \times C$ feature map,
which is then fed to the next convolutional layer to extract translation-shift invariant features.
In this case, the spatial regions of the feature maps are treated equally, which may contain noise or irrelevant information for the sound events.

In order to enhance the features from relevant time frames and frequency bands, 
temporal attention and spectral attention are presented in our work.
Different weights are applied to the time frames and frequency bands, 
which can guide the network to pay different attention to the temporal and spectral characteristics of environmental sounds.
The structures of the two attention mechanisms are depicted in Figure~\ref{fig1}.
Specifically, for the input feature map $\boldsymbol{U} \in \mathbb{R}^{T \times F \times C}$,
$1 \times 1 \times 1$ convolutional layers $f_{\mathrm{conv}}$ are employed to obtain the global feature maps across the channels,
\textit{i.e.} global temporal feature map $\boldsymbol{V}_{T} \in \mathbb{R}^{T \times F \times 1}$ and 
global spectral feature map $\boldsymbol{V}_{F} \in \mathbb{R}^{T \times F \times 1}$.
\begin{equation}
   \boldsymbol{V}_{T}=f_{\mathrm{conv}}\left(\boldsymbol{U} ; \theta_{\mathrm{T}}\right)
   \label{eq1}
\end{equation}
\begin{equation}
   \boldsymbol{V}_{F}=f_{\mathrm{conv}}\left(\boldsymbol{U} ; \theta_{\mathrm{F}}\right)
   \label{eq2}
\end{equation}
where $\theta_{\mathrm{T}}$ and $\theta_{\mathrm{F}}$ denote the model parameters of the convolutional layers in the temporal attention and the spectral attention, respectively.
The $1 \times 1$ filters are used to squeeze the number of channels to $1$,
which can learn channel-wise global information from the local feature map $\boldsymbol{U}$.
The extracted global temporal and spectral feature maps are then squeezed by the global average pooling $f_{\mathrm{GAP}}$ to obtain the time-wise activations $\boldsymbol{v}_{T} \in \mathbb{R}^{T\times 1\times 1}$ and the frequency-wise activations $\boldsymbol{v}_{F}\in \mathbb{R}^{1\times F \times1}$.
\begin{equation}
   \boldsymbol{v}_{T}=\sigma\left(f_{\mathrm{GAP}}\left(\boldsymbol{V}_{T}\right)\right)
   \label{eq3}
\end{equation}
\begin{equation}
   \boldsymbol{v}_{F}=\sigma\left(f_{\mathrm{GAP}}\left(\boldsymbol{V}_{F}\right)\right)
   \label{eq4}
\end{equation}
where $\sigma(\cdot)$ denotes the sigmoid function to limit the values in the range of $\left(0,1\right)$
and global average pooling is applied along the time axis and the frequency axis, respectively.
Thus, the time-wise feature map $\boldsymbol{U}_{T} \in \mathbb{R}^{T \times F \times C}$ and the frequency-wise feature map $\boldsymbol{U}_{F} \in \mathbb{R}^{T \times F \times C}$ can be obtained by rescaling $\boldsymbol{U}$
with the time-wise activations $\boldsymbol{v}_{T}$ and the frequency-wise activations $\boldsymbol{v}_{F}$.
\begin{equation}
   \label{eq5}
   \boldsymbol{U}_{T}=f_{\mathrm{scale}}\left(\boldsymbol{U} , \boldsymbol{v}_{T}\right)
\end{equation}
\begin{equation}
   \label{eq6}
   \boldsymbol{U}_{F}=f_{\mathrm{scale}}\left(\boldsymbol{U} , \boldsymbol{v}_{F}\right)
\end{equation}
where $f_{\mathrm{scale}}\left(\cdot , \cdot\right)$ refers to the multiplication between the feature map and the activations (\textit{i.e.} the time-wise activations $\boldsymbol{v}_{T}$ and the frequency-wise activations $\boldsymbol{v}_{F}$).

\begin{figure}[t] 
  \centering
  \subfigure[Temporal-spectral concatenation]{\label{h1}
  \begin{minipage}[c]{0.5\textwidth}
  \centering
  \includegraphics[width=2.8in]{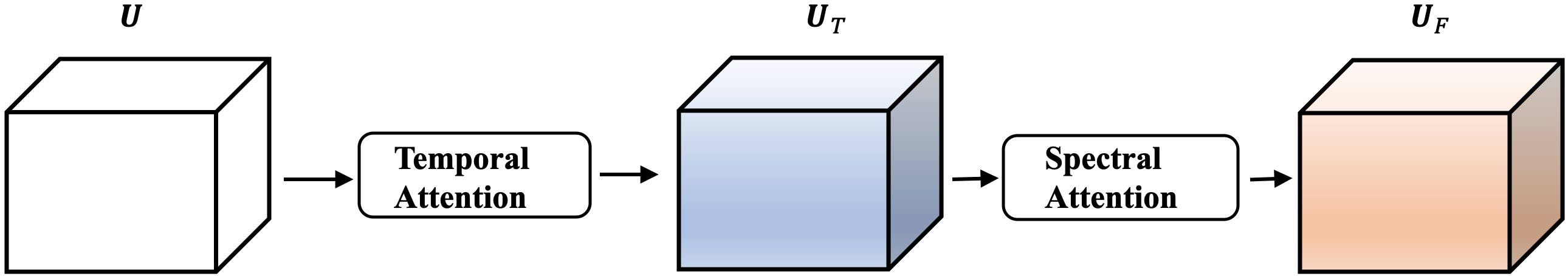}
  \vspace{0.2cm}
  \end{minipage}
  }
  \subfigure[Spectral-temporal concatenation]{\label{h2}
  \begin{minipage}[c]{0.5\textwidth}
  \centering
  \includegraphics[width=2.8in]{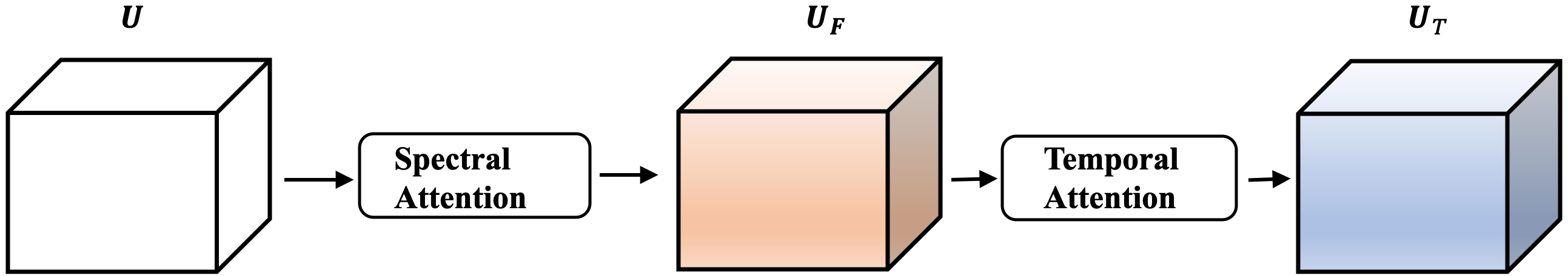}
  \vspace{0.2cm}
  \end{minipage}
  }
  \caption{The illustration of the concatenation of the temporal attention and the spectral attention.}
  \label{fig2}
\end{figure}



\begin{figure}[t]
   \begin{minipage}[c]{0.5\textwidth}
   \centering
   \includegraphics[width=2.8in]{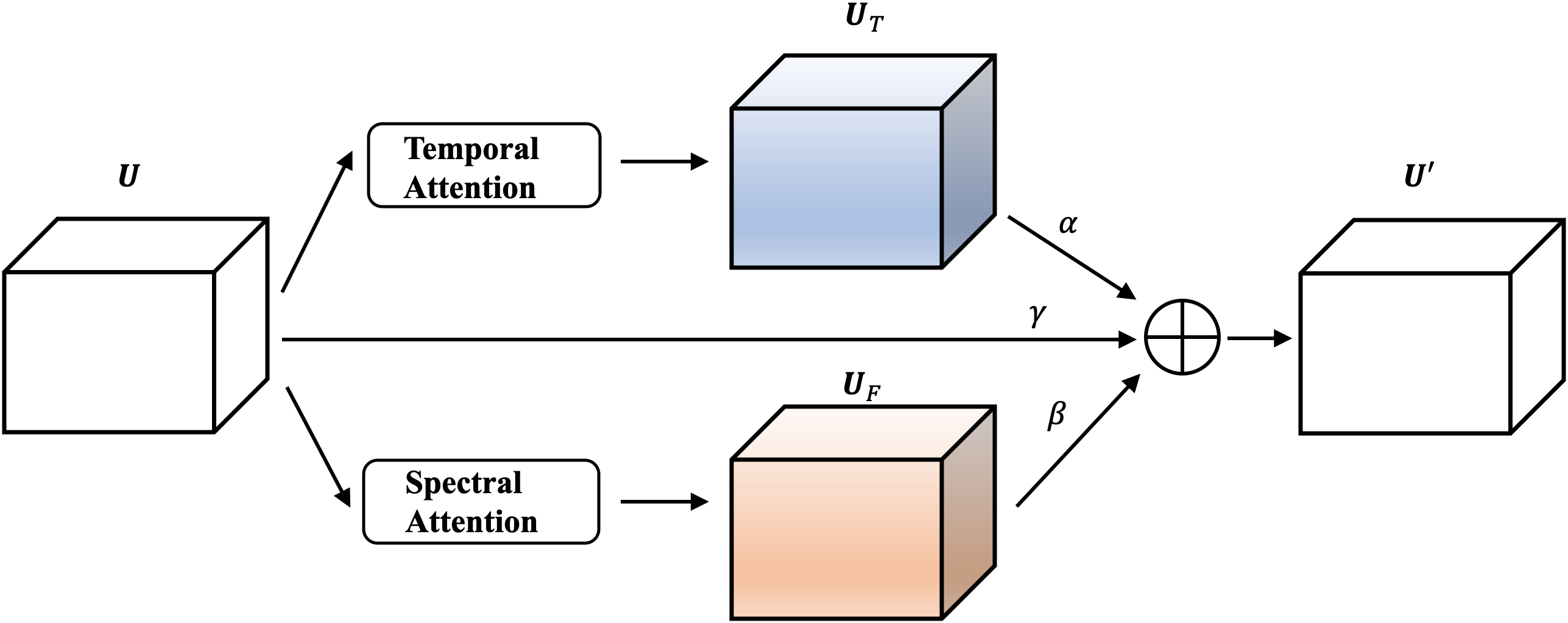}
   \caption{The illustration of the parallel temporal-spectral attention mechanism.}
   \label{fig3}
   \end{minipage}
\end{figure}

\subsection{Parallel Temporal-spectral Attention}
\label{sss}
One intuitive approach to obtain the temporal and spectral features simultaneously is the concatenation of the temporal attention and spectral attention, which is shown in Figure~\ref{fig2}.
However, there is a drawback in the approach of the concatenation, \textit{i.e.} the temporal attention and spectral attention may interfere with each other.
For example, when the temporal-spectral concatenation in Figure~\ref{fig2}(a) is applied, the activations of some noisy frames will be inhibited by the temporal attention,
while on the other hand, some slices of the noisy frames may be enhanced by the spectral attention.
Thus, the effect of the temporal attention is interfered by the spectral attention.

In order to alleviate the interference of the temporal attention and the spectral attention, 
a parallel temporal-spectral attention mechanism is proposed in our work.
As shown in Figure~\ref{fig3}, the temporal and spectral features are paid attention with two different branches without the propagation of information mutually.
In this case, each representation learning is focused on specific discriminative local regions rather than being spread evenly over the whole feature map,
which leads to a better robustness when a single branch is disturbed by the sections where no sound events appear in the acoustic environments.

To be specific, a summation of the three branches (\textit{i.e.} temporal attention, spectral attention and shortcut) are applied to obtain the final time-frequency features.
The summation is not with the same weights for the reason that different attention should attend the temporal and spectral characteristics.
Given coefficient as $\alpha$, $\beta$, $\gamma$, the time-frequency feature map $\boldsymbol{U}^{\prime} \in \mathbb{R}^{T \times F \times C}$ is calculated by the following formulas:
\begin{equation}
   \label{eq7}
   \boldsymbol{U}^{\prime} = \alpha\boldsymbol{U}_{T}+\beta\boldsymbol{U}_{F}+\gamma\boldsymbol{U}
\end{equation}
\begin{equation}
  \label{eq8}
  \alpha+\beta+\gamma=1
\end{equation}
where $\alpha$, $\beta$, $\gamma$ are the learnable parameters with the same initial value,
and the softmax function is applied to normalize them.
In this case, the network can adaptively pay different attention to the temporal characteristics and the spectral characteristics.



\section{Experiments}
Our method is evaluated on three ESC datasets (ESC-10 \cite{piczak2015esc}, ESC-50 \cite{piczak2015esc} and UrbanSound8k \cite{salamon2014dataset}) and two ASC datasets (the DCASE 2018 task1A dataset \cite{Mesaros2018_DCASE} and the DCASE 2019 task1A dataset \cite{Mesaros2018_DCASE}),
which are the commonly used datasets for ESC and ASC.
Log mel spectrograms are extracted from the audio signals as the input of the networks.
The experimental setups and results are detailed as follows.

\subsection{Datasets and Metrics}

\noindent\textbf{Datasets }The ESC-50 dataset \cite{piczak2015esc} comprises $50$ equally balanced classes of $2,000$ samples, and each sample is a monaural $5$s sound recorded with a sampling rate of $44.1$ kHz. 
The ESC-10 dataset \cite{piczak2015esc} is a selection of $10$ classes from the ESC-50 dataset.
Both datasets have been prearranged into $5$ folds for cross-validation.
The UrbanSound8k dataset \cite{salamon2014dataset} consists of $8,732$ audio clips summing up to $7.3$ hours of audio recordings. 
The maximum duration of audio clips is $4$s and there are $10$ unbalanced classes in the dataset. 
The original audio clips are recorded at different sample rates and the official $10$ folds are used in our experiments.
The DCASE 2018 task1A dataset \cite{Mesaros2018_DCASE} and the DCASE 2019 task1A dataset \cite{Mesaros2018_DCASE} contain $10$s segments, recorded at $48$kHz and spanning $10$ classes. 
The DCASE 2018 task1A dataset contains $8640$ segments in total with $6122$ segments for training and $2518$ segments for testing,
and the DCASE 2019 task1A dataset contains totally $13370$ segments with $9185$ segments for training and $4185$ for testing.

\noindent\textbf{Metrics }For all the used datasets, we use the accuracy of classification as the evaluation metric, which is one of the most commonly used metrics for audio classification \cite{virtanen2018computational}.

\subsection{Network Structures}
We set the experiments including a baseline model (CNN10 \cite{kong2019panns}) and ten comparison models under the same experimental setups to evaluate the proposed attention mechanism. 

\noindent\textbf{CNN10} CNN10 \cite{kong2019panns} consists of $4$ convolutional blocks with $64, 128, 256$ and $512$ output channels, respectively.
Each convolutional block contains $2$ convolutional layers with kernel size of $3 \times 3$, followed by downsampling with average pooling size of $2 \times 2$.
Batch normalization \cite{ioffe2015batch} and ReLU \cite{nair2010rectified} function are applied to all the convolutional layers.
Global pooling layer and two fully-connected layers are then applied, followed by a softmax nonlinearity for classification. 
See \cite{kong2019panns} for more details about CNN10.

\noindent\textbf{TS-CNN10} TS-CNN10 is our proposed model based on CNN10, where the parallel temporal-spectral attention mechanism is employed to each convolutional block.
All the other setups are the same as CNN10.
TS-CNN10-1, TS-CNN10-2, TS-CNN10-3 and TS-CNN10-4 are the variants of TS-CNN10, 
which only apply the parallel temporal-spectral attention mechanism to the 1st, 2nd, 3rd and 4th convolutional block, respectively.
TS-CNN10-fixed is another variant of TS-CNN10, where the parameters $\alpha$, $\beta$ and $\gamma$ in \eqref{eq7} are fixed to the same value ($\alpha=\beta=\gamma=0.33$).
T-CNN10 and S-CNN10 apply the temporal attention in Figure~\ref{fig1}(a) and the spectral attention in Figure~\ref{fig1}(b), respectively.
In addition, TS-CNN10-concat and ST-CNN10-concat apply the attention mechanisms of temporal-spectral concatenation in Figure~\ref{fig2}(a) and spectral-temporal concatenation in Figure~\ref{fig2}(b), respectively.

\subsection{Experimental Setups}
\noindent\textbf{Preprocessing }
All the raw audios are resampled to $44.1$kHz
and then fixed to the certain length by zero-padding or truncating 
(\textit{i.e.} $5$s for the ESC-10 and ESC-50, $4$s for the UrbanSound8k and $10$s for the DCASE 2018 task1A dataset and DCASE 2019 task1A dataset).
The short time Fourier transform (STFT) is then applied on the audio signals to calculate spectrograms, with a window size of $40$ms and a hop size of $20$ms. 
$40$ mel filter banks are applied on the spectrograms followed by a logarithmic operation to extract the log mel spectrograms.

\noindent\textbf{Training details }In the training phase, the Adam algorithm \cite{kingma2014adam} is employed as the optimizer with the default parameters. 
The model is trained end-to-end with the initial learning rate of $0.01$ and the exponential decay rate of $0.98$ for each $5$ iterations.
Parameters of the networks are learned using the categorical cross entropy loss. 
Batch size is set to 64 and training is terminated after $2000$ iterations. 
Data augmentation methods mixup \cite{zhang2017mixup} and Specaugment \cite{park2019specaugment} are applied in our experiments to prevent the system from over-fitting and improve the performance.

\begin{table}[t]\footnotesize
\begin{center}
\caption{Comparison of accuracy on the ESC-10, ESC-50 and UrbanSound8k (US8k) datasets} \label{t2}
\begin{tabular}{|c|c|c|c|}
  \hline
  \textbf{Model} & \textbf{ESC-10} & \textbf{ESC-50} & \textbf{US8k}\\
  \hline
  PiczakCNN \cite{piczak2015esc} & 80.5\% & 64.9\% & 73.0\%\\
  EnvNet-v2 \cite{tokozume2017learning} & 91.3\% & 84.9\% & 78.3\%\\
  SB-CNN \cite{salamon2017deep} & 91.7\% & 83.9\% & 83.7\%\\
  GTSC+TEO-GTSC \cite{agrawal2017novel} & - & 81.9\% & 88.0\%\\
  ConvRBM+FBEs \cite{sailor2017unsupervised} & - & 86.5\% & -\\
  ACRNN \cite{zhang2019attention} & 93.7\% & 86.1\% & -\\
  Multi-Stream CNN \cite{li2019} & 94.2\% & 84.0\% & -\\
  MelFB+LGTFB-EN-CNN \cite{park2020cnn} & 93.7\% & 88.1\% & 85.8\%\\
  \hline
  Human\cite{piczak2015environmental} & 95.7\% & 81.3\% & -\\
  \hline
  CNN10 \cite{kong2019panns} & 92.0\% & 85.2\% & 84.9\% \\
  \hline
  \textbf{T-CNN10 (ours)} & 94.5\% & 87.8\% & 87.2\%\\
  \textbf{S-CNN10 (ours)} & 94.0\% & 87.5\% & 87.8\%\\
  \textbf{TS-CNN10-1 (ours)} & 93.6\% & 86.9\% & 86.3\%\\
  \textbf{TS-CNN10-2 (ours)} & 93.8\% & 87.0\% & 87.0\%\\
  \textbf{TS-CNN10-3 (ours)} & 93.9\% & 86.9\% & 87.3\%\\
  \textbf{TS-CNN10-4 (ours)} & 94.9\% & 87.8\% & 87.5\%\\
  \textbf{TS-CNN10-concat (ours)} & 92.5\% & 85.6\% & 85.8\%\\
  \textbf{ST-CNN10-concat (ours)} & 93.0\% & 85.5\% & 86.1\%\\
  \textbf{TS-CNN10-fixed (ours)} & 95.0\% & 88.1\% & 88.3\%\\
  \textbf{TS-CNN10 (ours)} & \textbf{95.8\%} & \textbf{88.6\%} & \textbf{88.5\%}\\
  \hline
\end{tabular}
\end{center}
\end{table}

\begin{table}[t]\footnotesize
\begin{center}
\caption{Comparison of accuracy on the DCASE 2018 task1A dataset (DCASE2018 1A) and the DCASE 2019 task1A dataset (DCASE2019 1A)} \label{t3}
\begin{tabular}{|c|c|c|}
   \hline
   \textbf{Model} & \textbf{DCASE2018 1A} & \textbf{DCASE2019 1A} \\
   \hline
   Official baseline\cite{Mesaros2018_DCASE} & 59.7\% & 62.5\% \\
   \hline
   CNN10 \cite{kong2019panns} & 68.1\% & 69.6\% \\
   \hline
   \textbf{TS-CNN10 (ours)} & \textbf{68.7\%} & \textbf{70.6\%}\\
   \hline
\end{tabular}
\end{center}
\end{table}

\begin{figure*}[t]
\begin{minipage}[t]{.22\linewidth}
  \centering
\centerline{\epsfig{figure=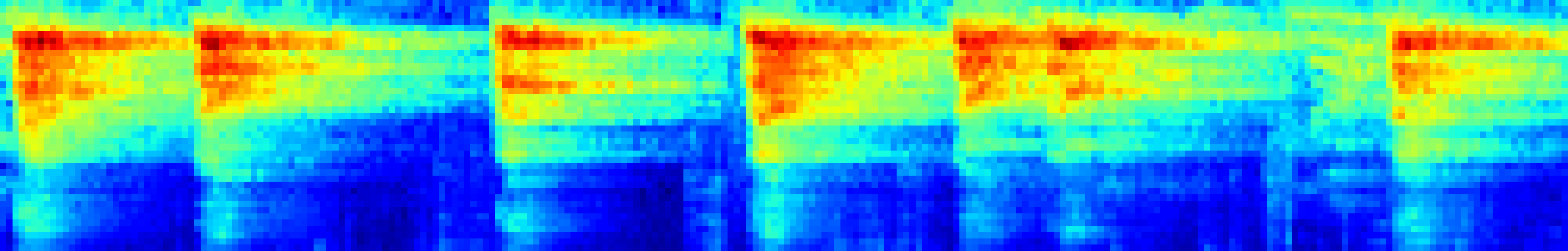,height=0.7cm}}
  \vspace{0.2cm}
\end{minipage}
\hfill
\begin{minipage}[t]{0.22\linewidth}
  \centering
\centerline{\epsfig{figure=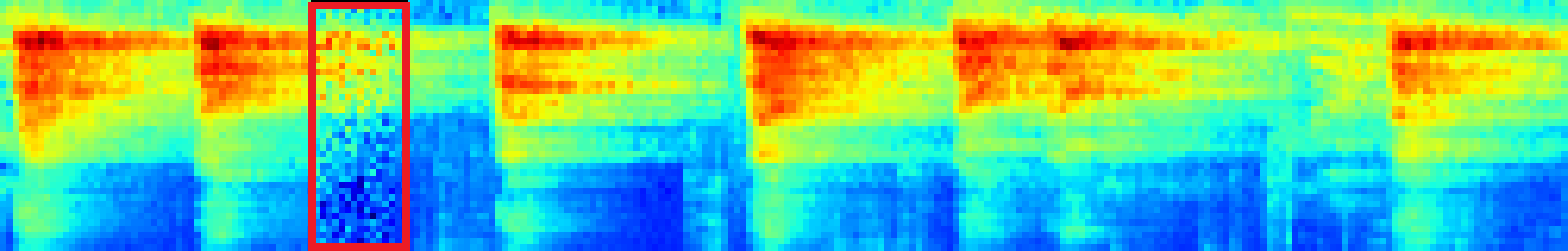,height=0.7cm}}
  \vspace{0.2cm}
\end{minipage}
\hfill
\begin{minipage}[t]{.22\linewidth}
  \centering
\centerline{\epsfig{figure=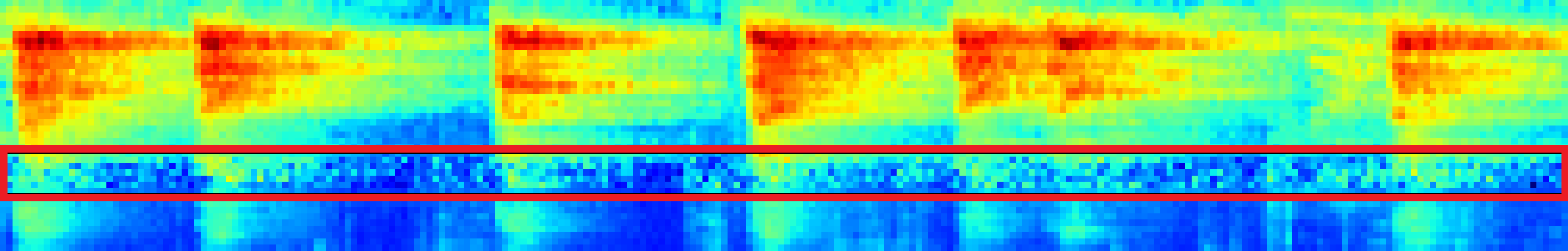,height=0.7cm}}
  \vspace{0.2cm}
\end{minipage}
\hfill
\begin{minipage}[t]{0.22\linewidth}
  \centering
\centerline{\epsfig{figure=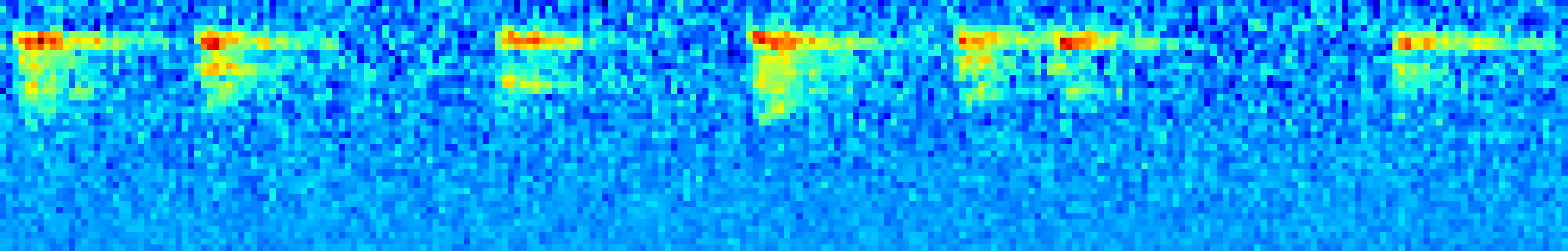,height=0.7cm}}
  \vspace{0.2cm}
\end{minipage}
\begin{minipage}[t]{.22\linewidth}
  \centering
\centerline{\epsfig{figure=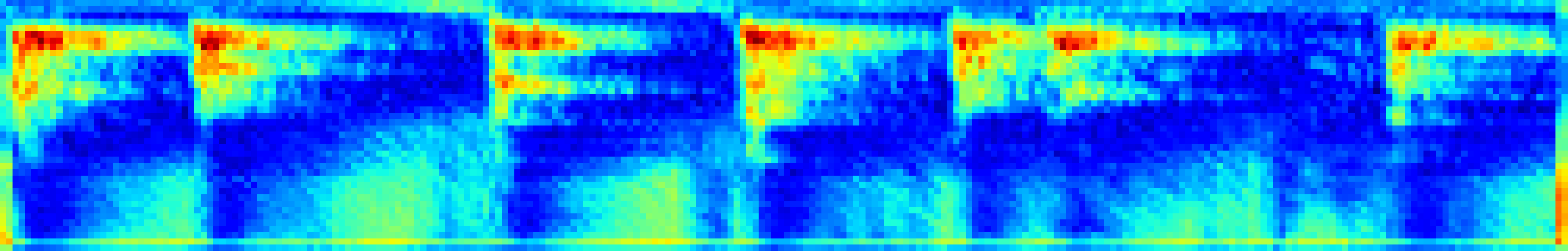,height=0.7cm}}
  \vspace{0.2cm}
\end{minipage}
\hfill
\begin{minipage}[t]{0.22\linewidth}
  \centering
\centerline{\epsfig{figure=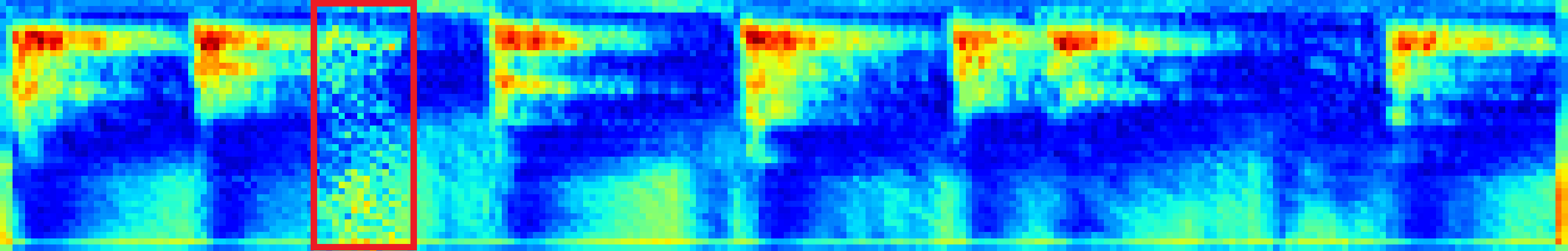,height=0.7cm}}
  \vspace{0.2cm}
\end{minipage}
\hfill
\begin{minipage}[t]{.22\linewidth}
  \centering
\centerline{\epsfig{figure=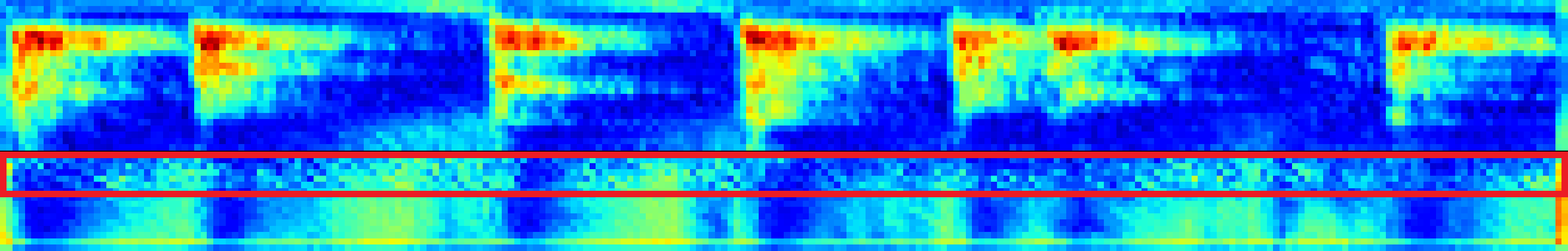,height=0.7cm}}
  \vspace{0.2cm}
\end{minipage}
\hfill
\begin{minipage}[t]{0.22\linewidth}
  \centering
\centerline{\epsfig{figure=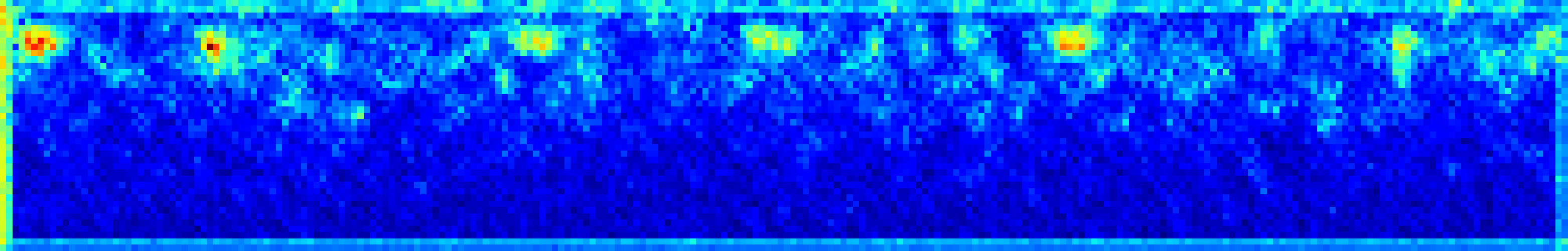,height=0.7cm}}
  \vspace{0.2cm}
\end{minipage}

\begin{minipage}[t]{.22\linewidth}
  \centering
\centerline{\epsfig{figure=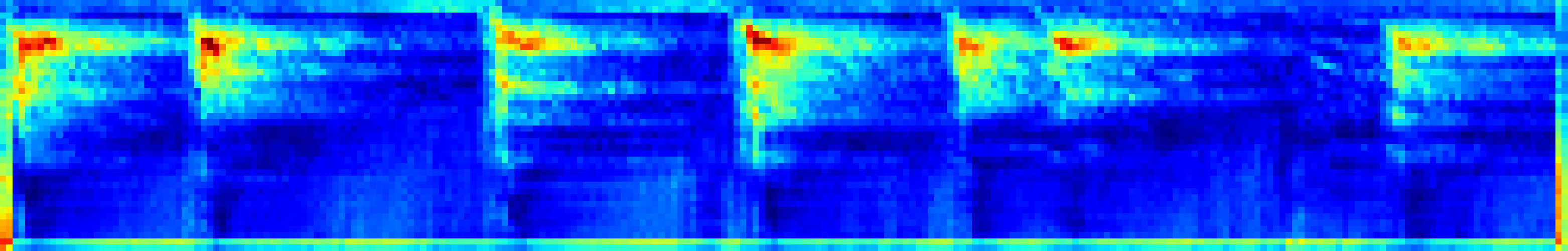,height=0.7cm}}
  \vspace{0.1cm}
  \centerline{(a) original audio}\medskip
\end{minipage}
\hfill
\begin{minipage}[t]{0.22\linewidth}
  \centering
\centerline{\epsfig{figure=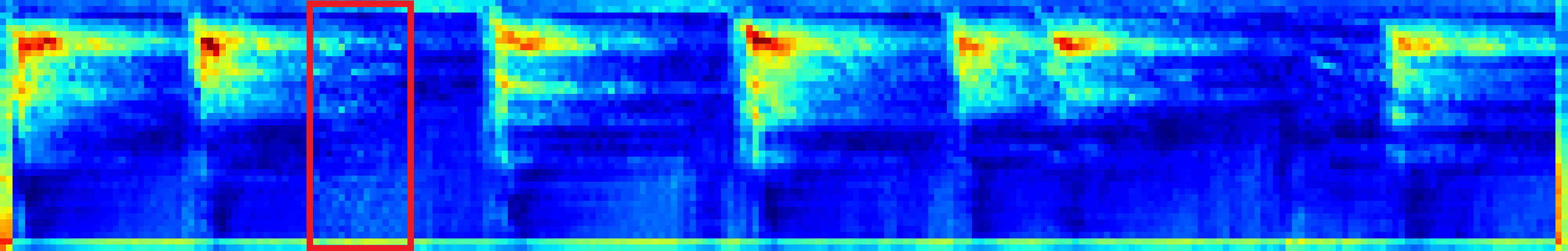,height=0.7cm}}
  \vspace{0.1cm}
  \centerline{(b) noisy audio \uppercase\expandafter{\romannumeral1}}\medskip
\end{minipage}
\hfill
\begin{minipage}[t]{.22\linewidth}
  \centering
\centerline{\epsfig{figure=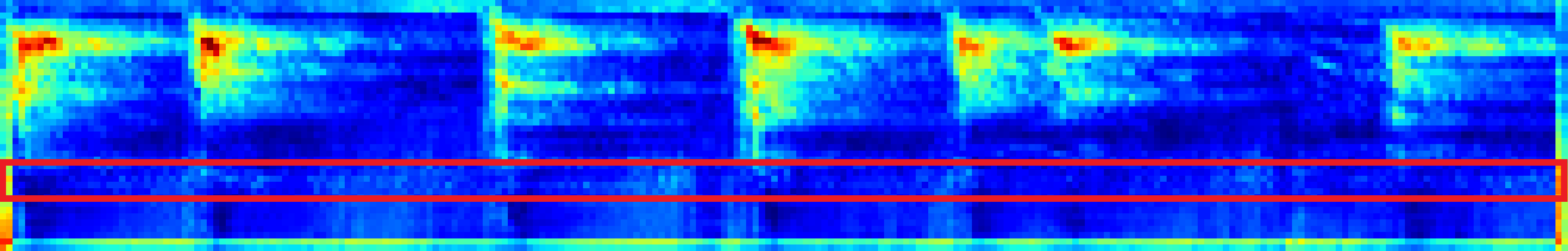,height=0.7cm}}
  \vspace{0.1cm}
  \centerline{(c) noisy audio \uppercase\expandafter{\romannumeral2}}\medskip
\end{minipage}
\hfill
\begin{minipage}[t]{0.22\linewidth}
  \centering
\centerline{\epsfig{figure=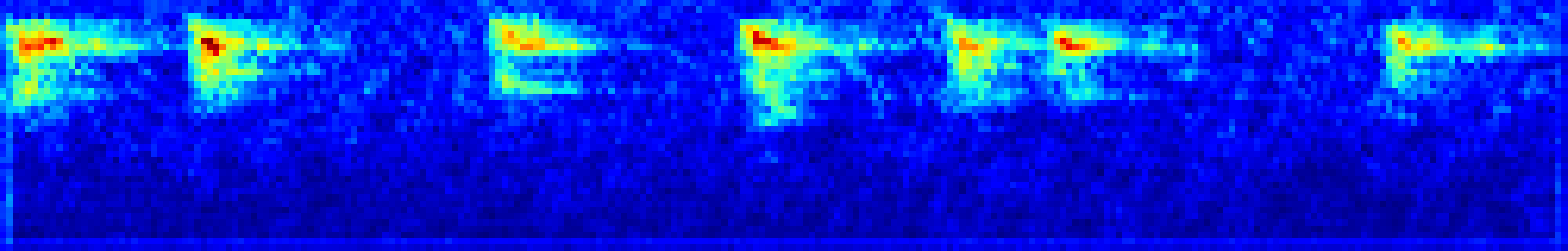,height=0.7cm}}
  \vspace{0.1cm}
  \centerline{(d) noisy audio \uppercase\expandafter{\romannumeral3}}\medskip
\end{minipage}
\caption{Visualization of four input log mel spectrograms (the 1st row) and the average feature maps of the first convolutional block in CNN10 (the 2nd row) and TS-CNN10 (the 3rd row).
(a) The original audio from the ESC-50 dataset. 
(b) Gaussian random noise is added into the $50$th$\sim$$64$th time frames (the red box). 
(c) Gaussian random noise is added into the $25$th$\sim$$30$th frequency bands (the red box). 
(d) Gaussian random noise is added into the whole audio with the SNR of $10$dB.}
\label{fig4}
\end{figure*}

\subsection{Experimental Results and Analysis}
Table~\ref{t2} demonstrates the performance of our proposed TS-CNN10 and other state-of-the-art methods on the ESC datasets (ESC-10 \cite{piczak2015esc}, ESC-50 \cite{piczak2015esc} and UrbanSound8k \cite{salamon2014dataset}).
Temporal attention was applied in \cite{zhang2019attention,li2019} to focus on the semantically relevant frames, which achieved higher accuracy than CNN models \cite{piczak2015esc,tokozume2017learning,salamon2017deep}.
However, the spectral characteristics of the environmental sounds were not considered in these methods.
Other methods \cite{agrawal2017novel,sailor2017unsupervised,park2020cnn} designed filter-bank learning approaches instead of feeding the log mel spectrograms to the networks, however, the deep time-frequency characteristics were hardly studied.
The results indicate that our proposed TS-CNN10 outperforms all the compared methods, which confirms the effectiveness of the proposed method in enhancing the features from relevant frames and frequency bands to obtain the discriminative sound representations.
It is worth mentioning that TS-CNN10 surpasses the performance of human being on both the ESC-10 dataset and the ESC-50 dataset.

We observe that both temporal attention (T-CNN10) and spectral attention (S-CNN10) improve the performance of ESC, and the combination of them (TS-CNN10) brings about more improvement. 
In addition, TS-CNN10-1, TS-CNN10-2, TS-CNN10-3, TS-CNN10-4 all perform better than CNN10, which shows that our proposed parallel temporal-spectral attention mechanism can be applied to any convolutional layer to enhance the features.
The parallel temporal-spectral attention used in the deeper layers shows more performance gain,
and a higher performance can be achieved when more layers apply the parallel temporal-spectral attention mechanism (TS-CNN10-fixed and TS-CNN10).
TS-CNN10 achieves higher accuracy than TS-CNN10-fixed for the reason that learnable parameters in \eqref{eq7} are set to adaptively pay different attention to the temporal and spectral characteristics.
Besides, TS-CNN10 outperforms TS-CNN10-concat and ST-CNN10-concat, which validates the advantage of the parallel structure to alleviate the interference of the temporal features and the spectral features.

Our method is also applied to another audio classification task (\textit{i.e.} ASC),
and evaluated on the DCASE 2018 task1A dataset \cite{Mesaros2018_DCASE} and the DCASE 2019 task1A dataset \cite{Mesaros2018_DCASE}.
As shown in Table~\ref{t3}, our attention mechanism can improve the performance of ASC as well.

To further test the robustness of TS-CNN10, three different types of noises (\textit{i.e.} Gaussian random, bus and tram) with different SNR (\textit{i.e.} $20$dB, $10$dB and $0$dB) are applied to the original audio in the ESC-50 dataset. 
CNN10 and TS-CNN10 are trained with the original data and then tested on the noisy data, with the results shown in Table~\ref{t4}.
TS-CNN10 shows better robustness to all the three types of noises, which is more obvious with the decrease of the SNR. 

In addition, visualization analyses are employed to show how our method enhances the time-frequency representations in a complex and dynamic acoustic environment.
As shown in Figure~\ref{fig4}, the input log mel spectrograms and the feature maps of CNN10 and TS-CNN10 are visualized. 
Figure~\ref{fig4}(a) is the original audio from the ESC-50 dataset.
It is clear that TS-CNN10 focuses more on the relevant time frames and frequency bands, 
and attenuates the less relevant information as compared with CNN10.
Figure~\ref{fig4}(b) and Figure~\ref{fig4}(c) show the noisy audio, which Gaussian random noises are added into several time frames and frequency bands.
CNN10 cannot deal with the noisy audio well and the feature maps are activated in the noisy sections.
While for our TS-CNN10, the feature maps in the noisy sections are much less activated. 
It is because the temporal-spectral attention is introduced in two different branches so that
computation for representation learning is focused on specific discriminative local regions rather than being spread across the whole feature map, which leads to a better robustness when a single branch is disturbed by the noisy sections.
To be specific, when random noise in several time frames is added, the spectral attention can mitigate the impact of the noisy sections by applying the global average pooling along the time axis. 
Similarly, we can explain the performance of TS-CNN10 in noisy frequency bands.
Moreover, Gaussian random noise is added to the whole audio with the SNR of $10$dB, which is shown in Figure~\ref{fig4}(d).
The result shows that the proposed TS-CNN10 can still obtain the time-frequency information of the sections where sound events appear. 
However, CNN10 can hardly capture the effective information and the feature map looks noisy.


\begin{table}[t]\footnotesize
\begin{center}
\caption{Comparison of accuracy on the ESC-50 dataset with different noise types and different SNR} \label{t4}
\begin{tabular}{|c|c|c|c|c|}
   \hline
   \textbf{Noise Type} & \textbf{SNR (dB)} & \textbf{CNN10} & \textbf{TS-CNN10}& \textbf{Gain}$^{\mathrm{\dagger}}$\\
   \hline
   No noise & - & 85.2\% & 88.6\% &3.4\% \\
   \hline
   \multirow{3}*{Gaussian random} & 20.0 & 83.5\% & 87.1\% &3.6\%\\
      &10.0 & 80.5\%& 85.9\% &5.4\%\\
      &0.0 & 71.2\%& 79.9\% &\textbf{8.7\%}\\ 
   \hline
   \multirow{3}*{Bus$^{\mathrm{\ast}}$} & 20.0& 82.3\%& 86.0\% &3.7\%\\
      &10.0 & 71.9\%& 76.9\% &5.0\%\\
      &0.0 & 58.3\%& 65.6\% &\textbf{7.3\%}\\ 
   \hline
   \multirow{3}*{Tram$^{\mathrm{\ast}}$} & 20.0& 78.5\%& 80.3\% &1.8\%\\
      &10.0 &62.7\%& 66.9\% &4.2\%\\
      &0.0 & 49.5\%& 53.9\% &\textbf{4.4\%}\\ 
   \hline
   \multicolumn{5}{p{220pt}}{$^{\mathrm{\dagger}}$ The accuracy gain of TS-CNN10 compared with CNN10. }\\
   \multicolumn{5}{p{220pt}}{$^{\mathrm{\ast}}$ The bus and tram noises are the clips randomly generated from the DCASE 2019 task1A dataset \cite{Mesaros2018_DCASE}.}\\
\end{tabular}
\end{center}
\vspace{-0.5cm}
\end{table}

\section{Conclusions}

A novel parallel temporal-spectral attention mechanism has been proposed to obtain the discriminative sound representations of environmental sounds.
Experimental results on the ESC and ASC tasks and visualization analyses validate the advantages of our method.
In future work, we would like to apply it to other audio classification tasks.

\end{document}